\documentclass[sigconf]{nimeart}
\AtBeginDocument{%
  }

\setcopyright{cc}
\copyrightyear{2025}
\acmYear{2025}
\acmDOI{}
\acmConference[NIME '25]{International Conference on New Interfaces for Musical Expression}{June 24--27,
  2025}{Canberra, Australia}
\acmISBN{}

\begin{document}

\title{LIMITER: A Gamified Interface for Harnessing Just Intonation Systems }

\author{Antonis Christou}
\email{antchris@media.mit.edu}
\orcid{0009-0003-3064-2100}
\affiliation{%
  \institution{Massachusetts Institute of Technology, Media Lab}
  \city{Cambridge}
  \state{Massachusets}
  \country{USA}
}


\renewcommand{\shortauthors}{Christou et al.}

\begin{abstract}
  This paper introduces LIMITER, a gamified digital musical instrument for harnessing and performing microtonal and justly intonated sounds. While microtonality in Western music remains a niche and esoteric system that can be difficult both to conceptualize and to perform with, LIMITER presents a novel, easy to pickup interface that utilizes color, geometric transformations, and game-like controls to create a simpler inlet into utilizing these sounds as a means of expression. We report on the background of the development of LIMITER, as well as explain the underlying musical and engineering systems that enable its function. Additionally, we offer a discussion and preliminary evaluation of the creativity-enhancing effects of the interface. 

\end{abstract}


\maketitle

\begin{figure}
  \centering
  \includegraphics[width=0.44\textwidth]{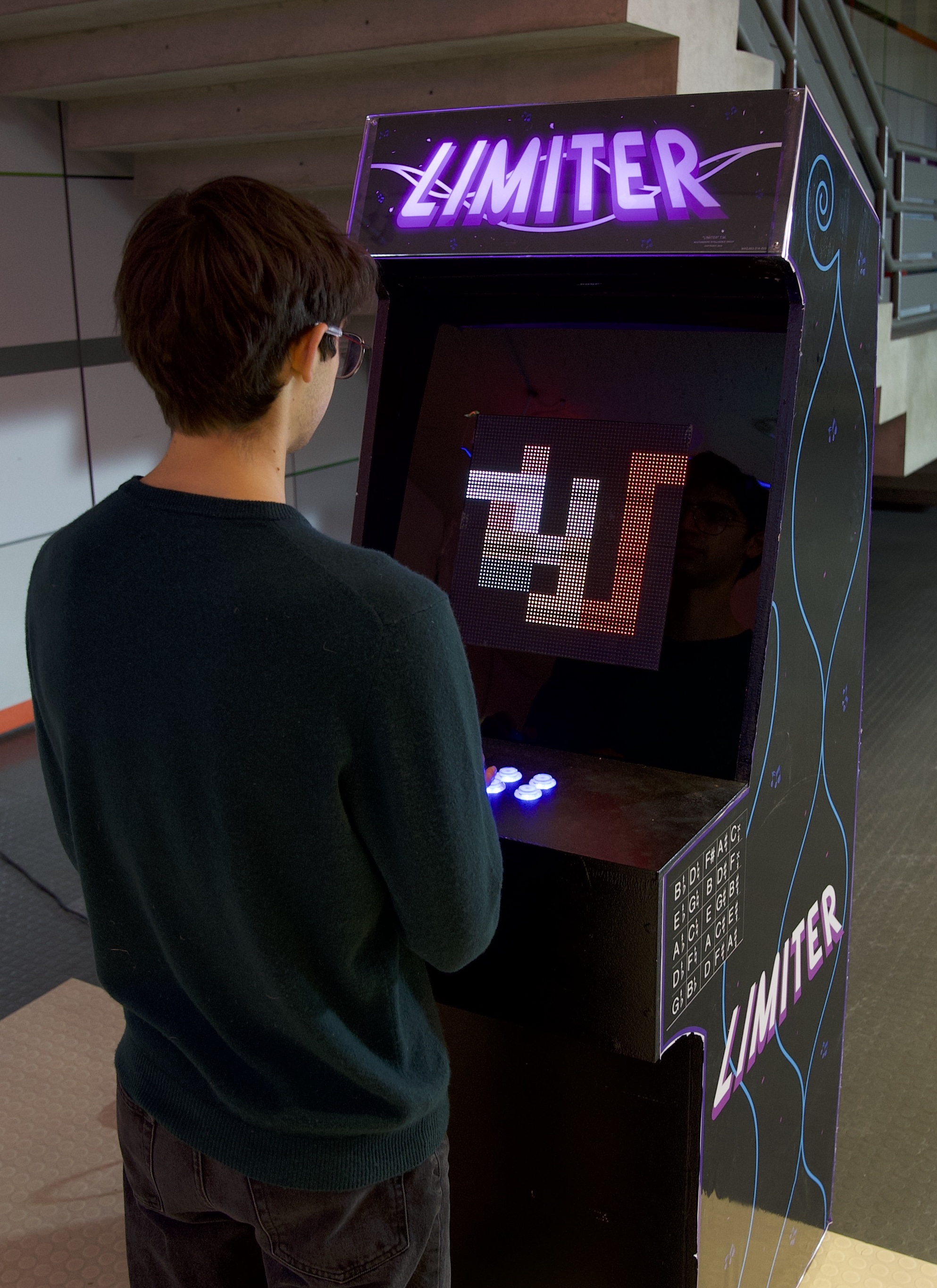}
  \caption{LIMITER pictured during performance.}
  \label{fig:perf}
\end{figure}

\section{Introduction}

Play is a fundamental element of human experience, and has been shown to have highly positive effects on development, mental health, and creativity \cite{ginsburg2007importance}. LIMITER is motivated, then, by a desire to incorporate game-like elements into a musical interface in order to capture the positive effects of play, particularly those on creativity, into music performance and composition.

Although microtonal and justly intonated music remain outside the realm of idiomatic and conventional Western popular music, with a few exceptions\footnote {an example of microtonality in popular music: \url{https://www.youtube.com/watch?v=HUGoUHKAGAE}}, we believe that harnessing these techniques offers the opportunity to discover new sounds and give audiences and performers new types of musical experiences.  

The primary problem with justly intonated music is that, while vocalists and non-fretted instrument performers often do it intuitively, asking musicians to perform specific, custom microtonal gestures with shifting tuning systems is a difficult feat that requires a large amount of training and rehearsal, and is largely restricted to experimental usage\footnote{a relatively well-known concert work that extensively features microtonality: \url{https://www.youtube.com/watch?v=VJkDH8o1-bs&t=14s}}. The advent of digital musical instruments (DMIs) in the second half of the twentieth century perhaps offers a solution to this problem, but many of these new interfaces face the fundamental limitation of lacking embodiment, with the literal sound-making device and the controller being separated and only abstractly connected, thus spurring the criticism of uninterpretability and unemotiveness from an audience's perspective \cite{Akito}.

LIMITER overcomes this problem by allowing the user to combine the experience of playing an 80s arcade machine with full justly intonated expression, offering the ability to swap tuning systems on the fly and play complexly tuned chords, but packaged in an interface that feels familiar and intuitive. With LIMITER, the musician uses the simple controls of a button and joystick to create these chords and communicate them with color and physical input in a way that abstracts the mathematical operations, but remains clear and highly expressive both to an audience and a performer. It offers a novel system of visual composition that allows amateurs and experienced musicians alike the ability to control these microtonal chords, thus synthesizing a gamified musical interface and a conduit for clearly visualizing alternate tuning systems into one creativity-enhancing system \footnote{\href{https://vimeo.com/1079127921}{Link to LIMITER video demo}}.

\section{Background}

LIMITER's uniqueness stems from its combination of both microtonal sound synthesis and a gamified interface with a visual, color-based system of performance and composition. While each of these have some precedent in isolation, this novel combination is what allows for LIMITER's ease of use and depth of expression. The remainder of this section is devoted to a brief discussion of prior NIMEs that are relevant to each of LIMITER's subsystems.

\subsection{Microtonal \& Justly Intonated NIMEs}

Numerous acoustic musical instruments from various cultures and time periods offer the potential for microtonal expression, ranging from non-fretted string instruments like the violin to the voice to the bağlama from Turkey and Greece \cite{tamer2013microtonal}. Many digital musical instruments, however, have  coalesced around using twelve tone equal temperament (12TET) due to the simple fact that the now ubiquitous MIDI standard bakes this system in to its protocol (although it is important to note that MIDI 2.0 does enable the use of some limited microtonal pitch bends) \cite{hamelberg2023applications}.

Despite this, over the course of NIME's proceedings from 2001 to the present there have been a number of papers that presented interfaces for microtonal or justly intonated expression. In one \cite{robertson2018harmonic}, the performer uses gestures and the excitation of various physical surfaces to control microtonal sound. Another system \cite{hayward2015vine} presented a physical interface for manipulating a harmonic tuning lattice in a way that bares some similarity to LIMITER's sound synthesis technique. Other systems focus on bringing the possibility of microtonal gestures and expression to woodwind instruments, with one \cite{chin2021hyper} focusing on breath control and another highlighting the potential for making 3D printable microtonally tuned flutes \cite{ritz20153d}. These instruments, despite their innovations, however, are generally either more fixed in the specific microtonal sounds they can create or remain difficult to understand from an audience perspective.

\subsection{Gamified Music and Compositions}

One of the core characteristics of play has been defined as being intrinsically motivated and lacking a clear goal \cite{rubin2018play}. In this way, the mere definition of play counters the traditional paradigm of "serious" and traditional Western art music that centers around extremely clearly defined parameters of instrumentation, form, and when and how the piece starts and stops. Even so, research has consistently shown the value of play towards enhancing creativity, among other positive effects, thus creating space for "playful" or gamified NIMEs \cite{Bateson_Martin_2013}.

Reflecting the above definition of play, the first "gamified" pieces of music can in some sense be seen as the non-deterministic and aleotoric works of the mid 20th century, such as Terry Riley's "In C"\footnote{\url{https://www.youtube.com/watch?v=yNi0bukYRnA}} and Karlheinz Stockhausen's  "Klavierstück XI"\footnote{\url{https://www.youtube.com/watch?v=mMDdihXI98A}}. These pieces give the performer a notable and unprecedented amount of freedom to choose how the music unfolds, enabling a sort of playful form of expression even in the usually strict world of Western classical music. 

Over the course of NIME and other related conferences' proceedings, a number of pieces have advertised themselves more directly as being gamified or incorporating game-like elements. Notable prior work harnesses standard game-like elements of levels \cite{michalakos2020icarus}, multiplayer collaboration or competition, and sonic feedback aimed at shaping the behavior of the player-performer(s). Existing research mostly has occurred in the domains of dance \cite{nime23-installations-dance}, or augmented acoustic instruments like the drumset \cite{michalakos2016pathfinder} or violin \cite{nime23-music-splt}. While this research is notable for allowing new forms of expression that incorporate playful elements, no existing work aims to synthesize the known positive effects of play into a creative interface that goes beyond the realization of a single piece.

\subsection{Visual Composition Systems and Instruments}

Finally, LIMITER uses a visual composition system as its primary way of controlling and communicating expressive justly intonated sound synthesis to an audience in a way that does not need or require extensive background in the theory behind it. LIMITER expands past efforts at visualizing tuning lattices\footnote{\url{https://www.youtube.com/watch?v=CSL_Axohw94}} by formalizing the previously fully digital presentation into a physical interface, and replacing the abstract mathematical notation with a clear, more easily understood system that revolves around color. 

Additionally, LIMITER is an example of a grid-based interface - a common paradigm for NIMEs ever since the 2007 Yamaha Tenori-on displayed the extensive expressive potential of the technique \cite{nishibori2006tenori}. The key insight of grid interfaces lies in their ability to quickly and intuitively express the possible field of musical actions via color and shape to the user \cite{Rossmy2021grid}, thus making them a perfect candidate for communicating the esoteric tuning information of LIMITER in a way that is easy to understand.

\begin{figure}
  \centering
  \includegraphics[width=0.44\textwidth]{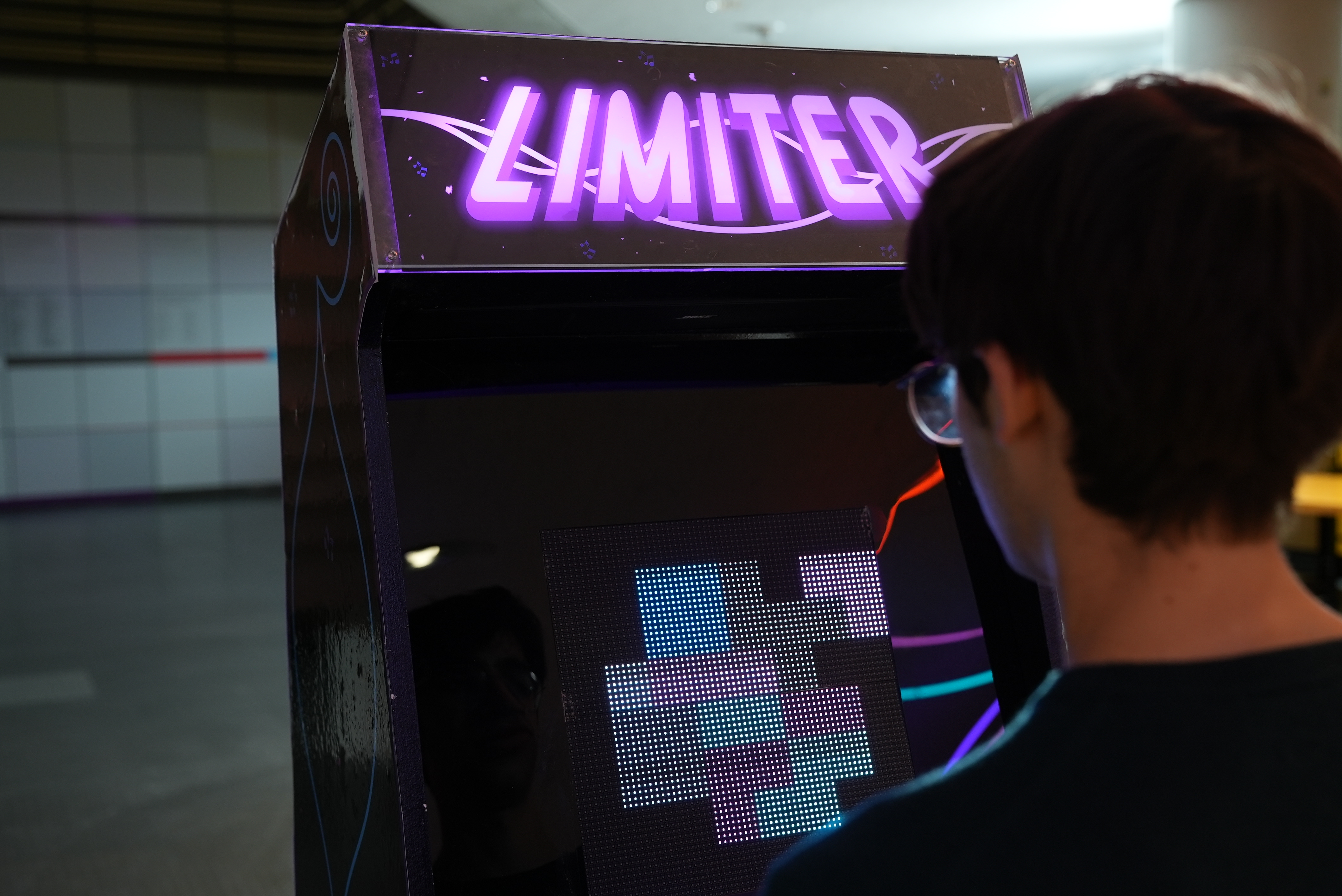}
  \caption{The illuminated marquee and screen captured during gameplay.}
  \label{fig:marquee}
\end{figure}

\section{Tuning}
LIMITER's unusual tuning system justifies a brief explanation of the history and mathematics that underlie it. The following section will, therefore, outline the theory behind its musical output, even though a full understanding is not required to be a skilled performer. 

\subsection{Microtonality and Just Intonation Systems}

The Western musical world has for hundreds of years settled on using
twelve-tone equal temperament as its standard for tuning and performance. Despite the existence of many other possible tuning systems and numerous other cultural traditions that utilize what the Western world perceives as microtonal scale degrees, Western music now largely relies on this division of the octave into twelve equal parts. Initially invented in 16th century China, this system quickly spread throughout Europe and allowed for enharmonic modulations and the ability to play in any given key without distant ones being grievously out of tune \cite{partch1949genesis}.

As is now widely known, however, this system makes every key and most intervals slightly out of tune from their "proper" harmonic ratios that derive from the harmonic series. The major third in this system, for example, lies 14 cents higher than what it should be, and the perfect fifth two cents lower. LIMITER instead makes use of just intonation systems based upon the proper harmonic ratios of nature as a means of correcting this discrepancy, thus enabling the exploration of new harmonic colors and tonal soundscapes.

\[
\left(\sqrt[12]{2}\right)^{12} = 2, \quad
\left(\sqrt[12]{2}\right)^{7} \approx 1.4983 \neq \frac{3}{2}, \quad
\left(\sqrt[12]{2}\right)^{4} \approx 1.2599 \neq \frac{5}{4} 
\]
\begin{center}
    \textbf{Figure 4: A brief example showing the deviation from 12TET to Just Intonation for the Perfect Fifth and Major Third.}
\end{center}

\subsection{The Tuning Lattice}

The remainder of this section is devoted to explaining LIMITER's specific system of five and seven limit just intonation, all based around the concept of the harmonic tuning lattice, as introduced by Leonard Euler in the 18th century \cite{euler1739tentamen}. This alternative way of arranging notes places them not in a sequential scale as is conventional, but rather in a multidimensional grid. In five limit just intonation, one multiplies a given center pitch by prime numbers less than or equal to five. Following past convention\footnote{\url{ https://www.youtube.com/watch?v=ZJfAVSVgaSI}}, we define the vertical axis as moving by justly tuned fifths (a ratio of 3/2), and the horizontal axis as moving by justly tuned thirds (a ratio of 5/4), creating the lattice seen in Figure 3.

\begin{figure}
  \centering
  \includegraphics[width=0.44\textwidth]{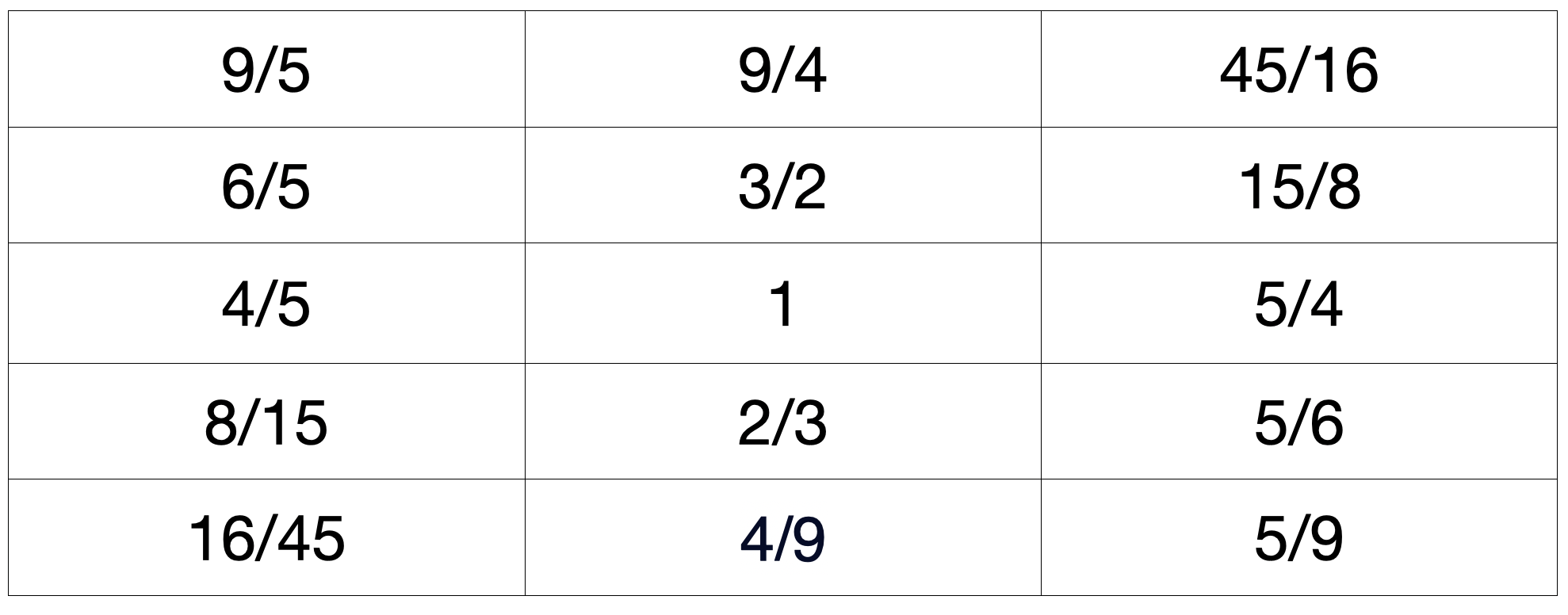}
  \caption{Tuning lattice of ratios created by multiplying and dividing by 3/2 in the vertical axis and 5/4 in the horizontal.}
  \label{fig:marquee}
\end{figure}

\setcounter{figure}{4}

We then can multiply each numerator or denominator by a power of two in order to bring the ratios into the range of one octave, or in other words, a ratio between one and two. These ratios can then be applied to create actual frequency values, centered in this case around the note of A4. In order to signify deviations from 12TET tuning, we use Helmholtz-Ellis notation \cite{sabat2009helm}, a system of visually conveying alterations of one or more Syntonic commas, a distance of approximately 21.5 cents.

The second tuning system used in LIMITER is seven limit just intonation, specifically based upon a simplified form of the one used by the American experimental composer La Monte Young in his seminal work for solo piano, "the Well Tuned Piano". This system utilizes both the justly tuned fifth of 3/2 as well as the "septimal seventh" ratio of 7/4, which at 969 cents is substantially flatter than the usual 12TET minor 7th, thus allowing for new, uniquely tuned chords.

\begin{figure}
  \centering
  \includegraphics[width=0.44\textwidth]{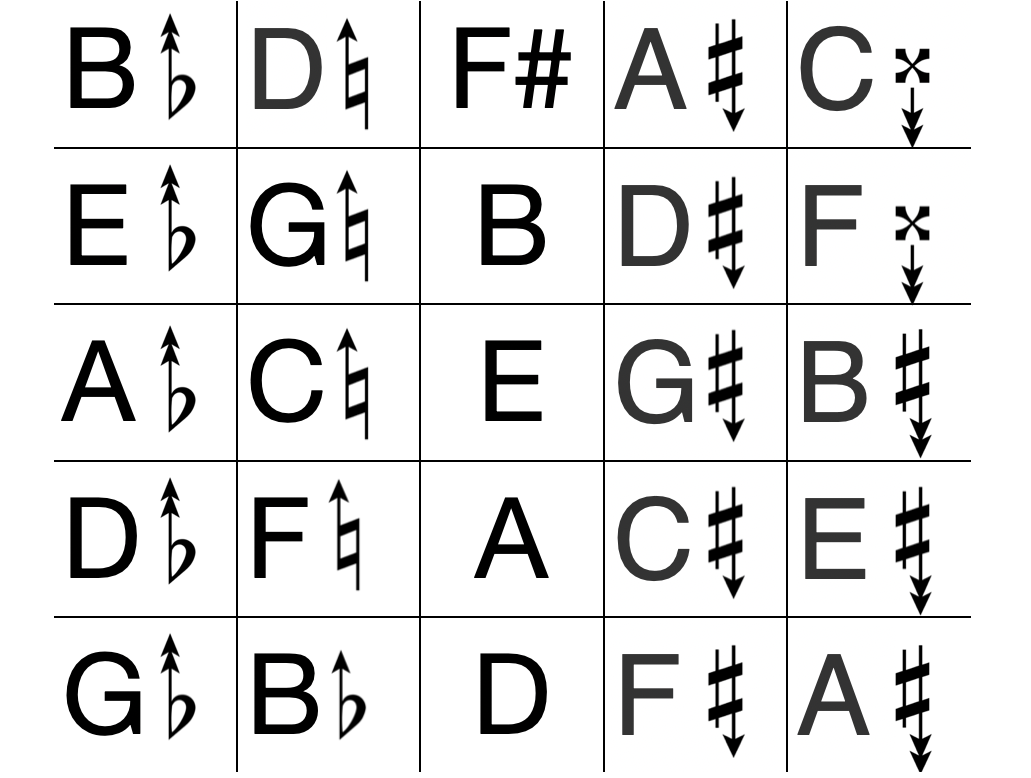}
  \caption{The five limit tuning lattice used in LIMITER that also is featured on the side art of the cabinet itself.}
  \label{fig:marquee}
\end{figure}

\begin{align*}
\mathbf{12TET\ Minor\ 7th:} \quad &\left(\sqrt[12]{2}\right)^{10} \times 440 = 783.99 \text{ Hz}, \\
&1200 \times \log_2\left(\frac{783.99}{440}\right) = \mathbf{999.96} \text{ cents}, \\
\mathbf{Septimal\ 7th:} \quad &\frac{7}{4} \times 440 = 770 \text{ Hz}, \\
&1200 \times \log_2\left(\frac{770}{440}\right) = \mathbf{968.88} \text{ cents}
\end{align*}

\begin{center}
    \textbf{Figure 6: Showing the large deviation of the septimal 7th compared to the 12TET minor 7th.}
\end{center}

\setcounter{figure}{6}

\section{LIMITER}

The name of the instrument derives from the various limit-based just intonation systems that are used in its musical output. Performance takes the form of the user manipulating a four-way joystick and eight buttons to draw and transform a series of shapes on the screen that correspond to justly tuned chords. Each button offers a different geometric transformation operation of the current chord, or move in the "game" that is performing with LIMITER. The basic "rules" are the following:

\begin{enumerate}
   \item {The screen is divided into a grid that corresponds to coordinates indexing into a tuning lattice. }
   \item Play consists of drawing a series of shapes on the screen. Each shape must have exactly four blocks, implying a four-note chord.
   \item The very first chord can be placed anywhere, so long as the blocks are contiguous vertically or horizontally (diagonals excluded). 
   \item After the first chord, the user can draw any new shape, with the restriction that it must share at least one overlapping block with any previous one, with a few exceptions that will be discussed in the section on input devices. 
   \item Each shape is assigned its own color, with the colors cycling as they eventually repeat. 
\end{enumerate}

These rules exist to constrain the possible field of musical possibilities, both to make playing the instrument less overwhelming and also to guide to the user towards certain sonic patterns. These patterns each have rough analogues in traditional Western music theory, but roughly center around the principle of generally guiding the player to connect and voice lead chords smoothly, usually with at least one shared note (in other words, a pivot tone modulation). These constraints help ensure that the absence of a fixed tonal center does not sound especially jarring, and encourages the player to hover around various "pools" of tonality, while perhaps even briefly establishing whispers of conventional functional harmony.

\begin{figure}
  \centering
  \includegraphics[width=0.44\textwidth]{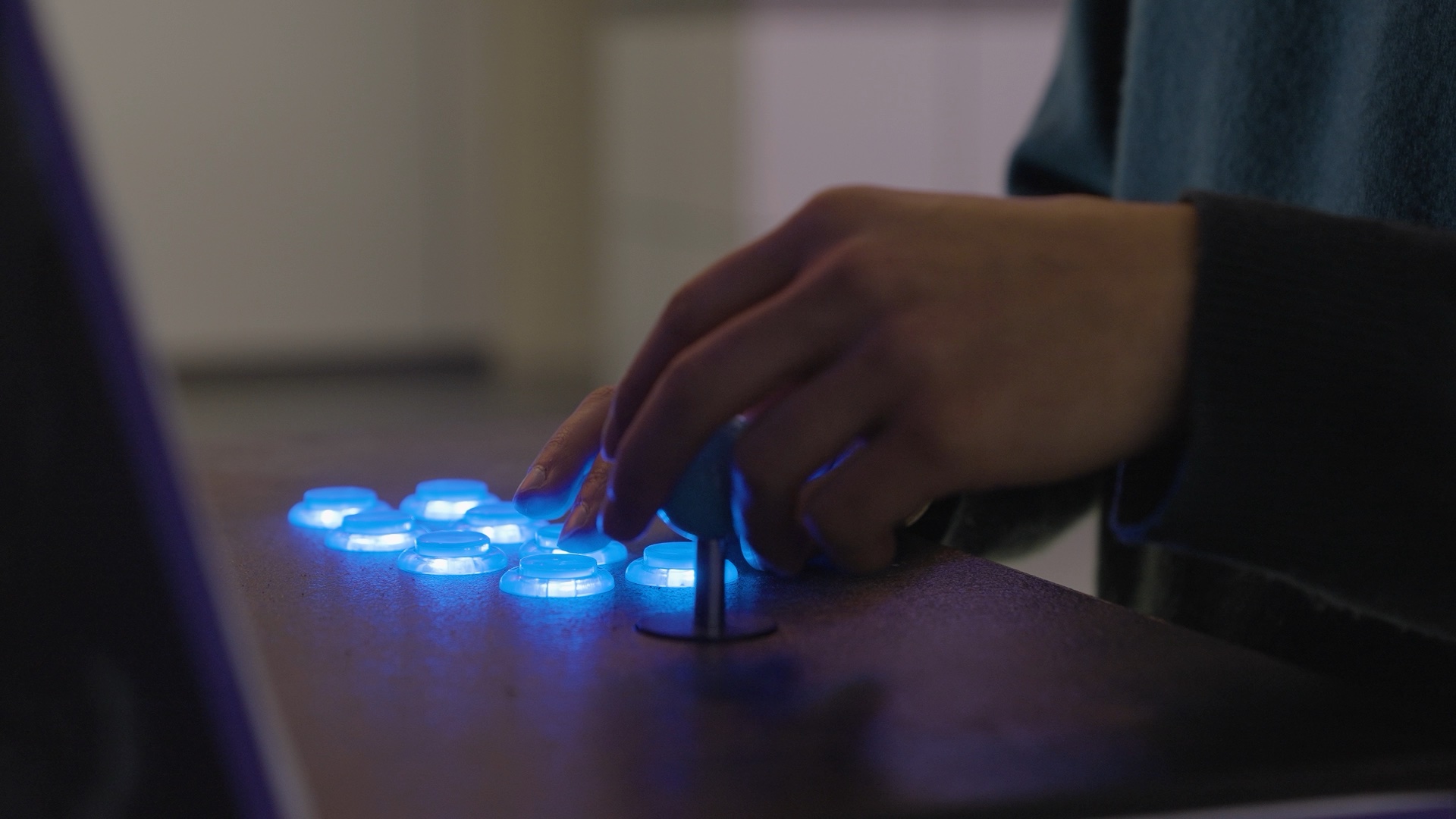}
  \caption{The joystick and buttons used for musical and game control.}
  \label{fig:hands}
\end{figure}

\subsection{Physical Specifications and Construction}

LIMITER physically and aesthetically resembles a classic 80s arcade machine. We modeled the instrument in CAD-based software and constructed a number of smaller scale models before cutting and assembling the final full-sized version using a CNC machine and table saws. LIMITER is made from a mixture of MDF and plywood, chosen for its historical authenticity and ease of finishing. The instrument stands at approximately two feet wide and six feet tall and is mounted on locking casters for ease of transportation.

The electronics are housed inside the instrument, secured directly under the control panel where the joystick and buttons are attached. LIMITER uses two microcontrollers and a Raspberry Pi 4 Model B working in tandem, communicating via a simple UART connection that was chosen for its speed and ease of use. A number of fabrication processes were used, ranging from PCB board milling to laser cutting for the acrylic panels to 3D printing for the custom-made brackets for mounting the speaker and screen.

\begin{figure}
  \centering
  \includegraphics[width=0.44\textwidth]{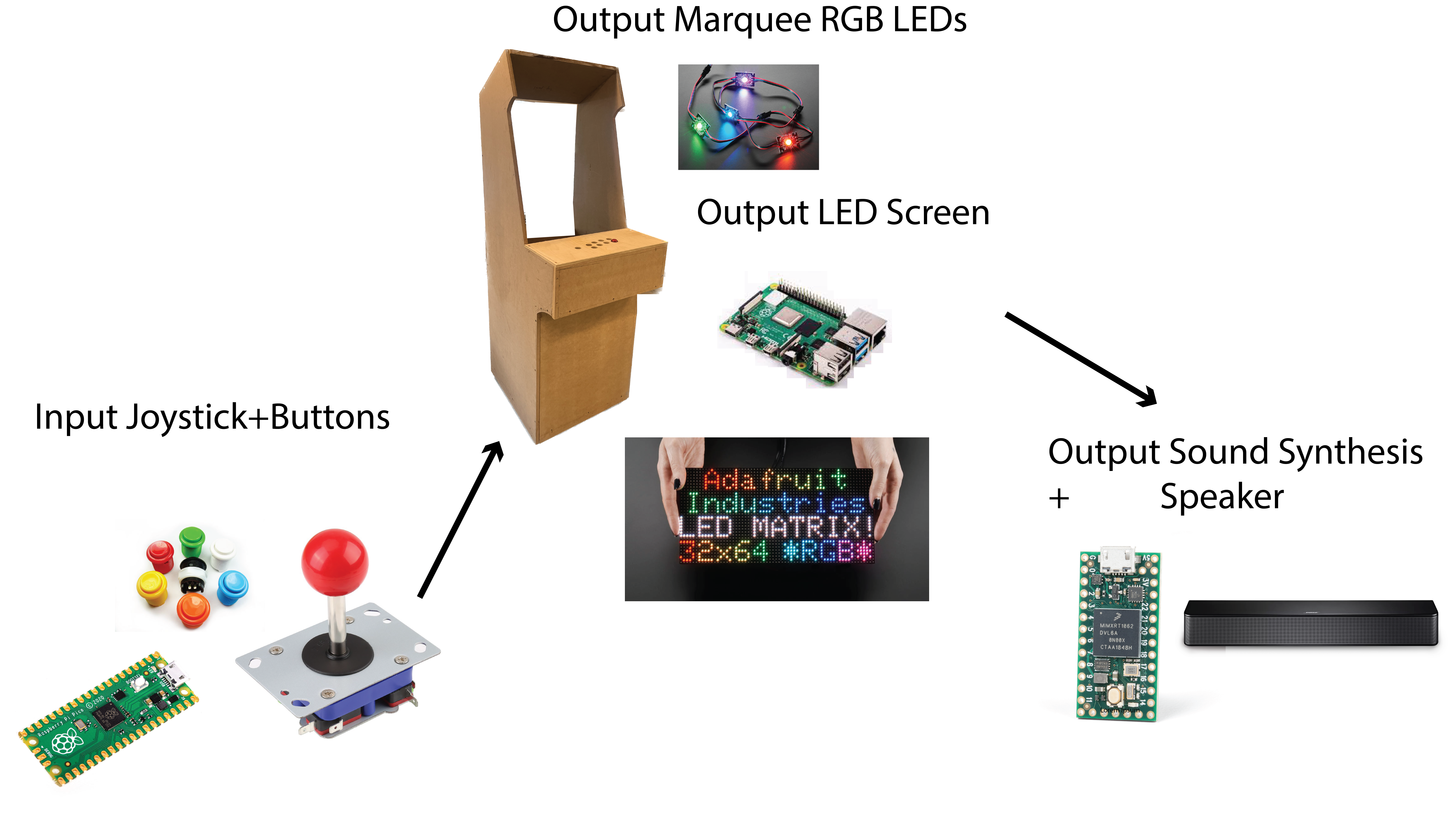}
  \caption{The full systems diagram used in LIMITER.}
  \label{fig:systemdiagram}
\end{figure}

\subsection{Input Devices}

The input devices take the form of a single joystick and eight buttons, which are controlled by a Raspberry Pi Pico running on a custom made PCB.

The joystick is used to move the player's cursor around the screen, allowing them to control where they place and draw shapes. The buttons each contain an LED within, and represent a particular musical effect or geometric transformation:

\begin{enumerate}
   \item {DRAW: illuminates the block that the cursor is currently hovering over, thus marking it as part of the current prospective chord.}
   \item SONIFY: takes the coordinates of the four blocks that are illuminated and plays them out via the speakers. Does nothing if four blocks are not yet drawn or if the current shape is invalid in some way. 
   \item SHIFT: changes mode from highlighting the current cursor location to highlighting the most previously sonified shape. The user can then slide the entire shape to anywhere on the screen, without the usual restriction of having to share one block with a previous chord. Does nothing if at least one shape has not yet been sonified. This operation corresponds musically to the "planing" technique used by Impressionistic composers to transport the user to nonfunctionally related key centers \cite{duffelDebussy}. 
   \item MIRROR: mirrors the previous shape over the center line of the screen.
   \item ROTATE: rotates the previous shape 90 degrees around a pivot block.
   \item DELETE: deletes the shape that is currently being drawn, allowing the user to redraw if they have made a mistake or wish to change the next chord. 
   \item CHANGE TUNING: flashes the screen in a burst of white, removing all blocks but the most previous one as the tuning lattice dynamically swaps between five and seven limit just intonation, giving the performer a different set of harmonic tools.  
   \item END GAME: turns the sound synthesis engine off and slowly fades the screen away, ending the performance. 

\end{enumerate}

\begin{figure}
  \centering
  \includegraphics[width=0.3\textwidth]{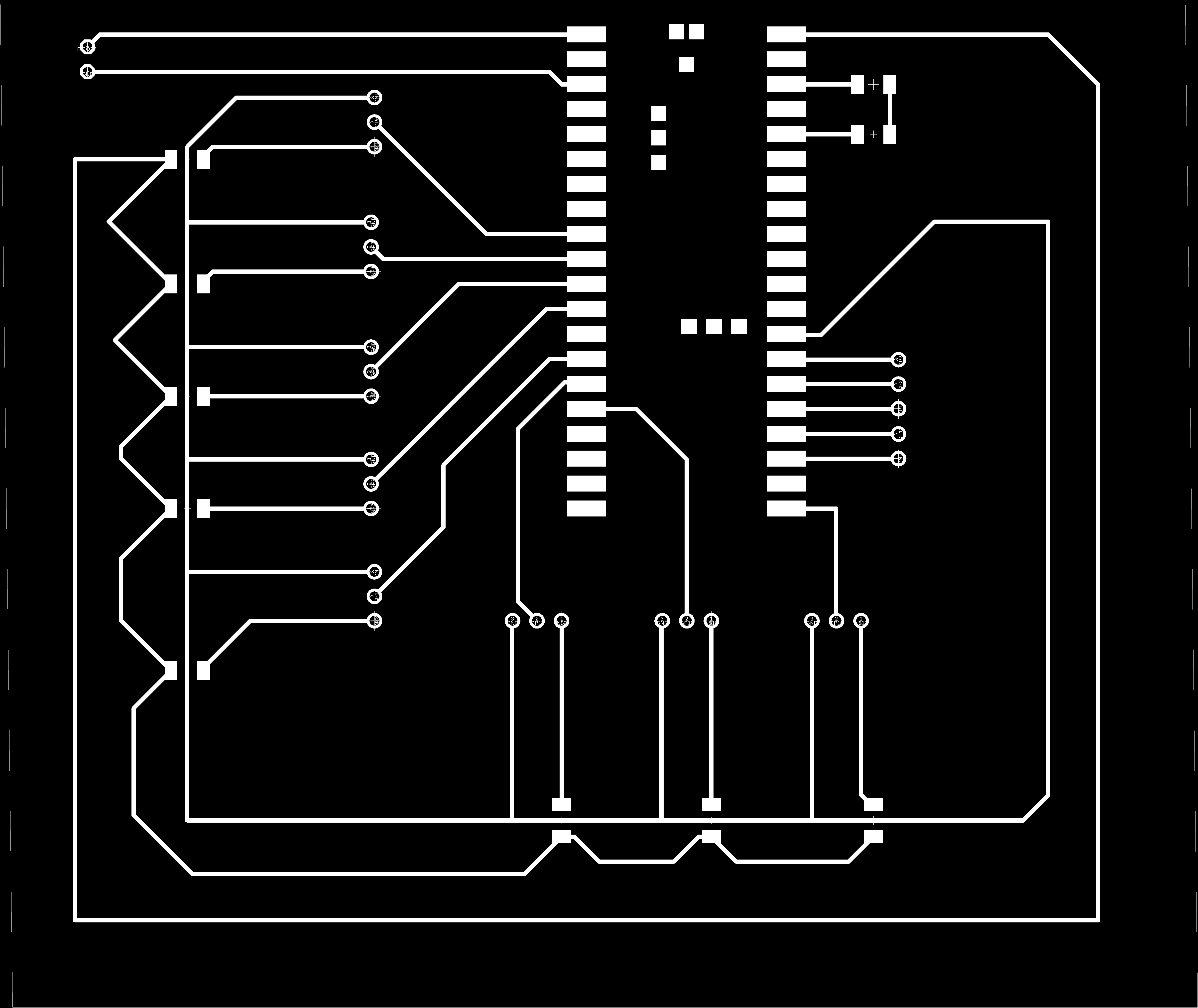}
  \caption{An early prototype PCB design for the input devices.}
  \label{fig:pcb}
\end{figure}

\subsection{Output Devices \& Sound Synthesis}

The visual and audio system is primarily controlled by a Raspberry Pi 4 Model B in conjunction with a Teensy 4.0 with audio shield. The Raspberry Pi acts as the main controller, coordinating communications between the input Pi Pico and the Teensy and driving the LED system. The screen of the arcade machine consists of a series of chained LED matrices, while the lights that illuminate the marquee are a NeoPixel individually addressable LED strip. Sound is played back with an embedded Bose Solo 5 Soundbar which operates in tandem with the Teensy. A design challenge that took particular care was designing a structurally sound 3D printed bracket that both exactly fit the dimensions of the arcade cabinet and the speaker, but also was secure enough to hold the speaker's weight. This took several iterations before the result was both strong and entirely precise. 

\begin{figure}
  \centering
  \includegraphics[width=0.3\textwidth]{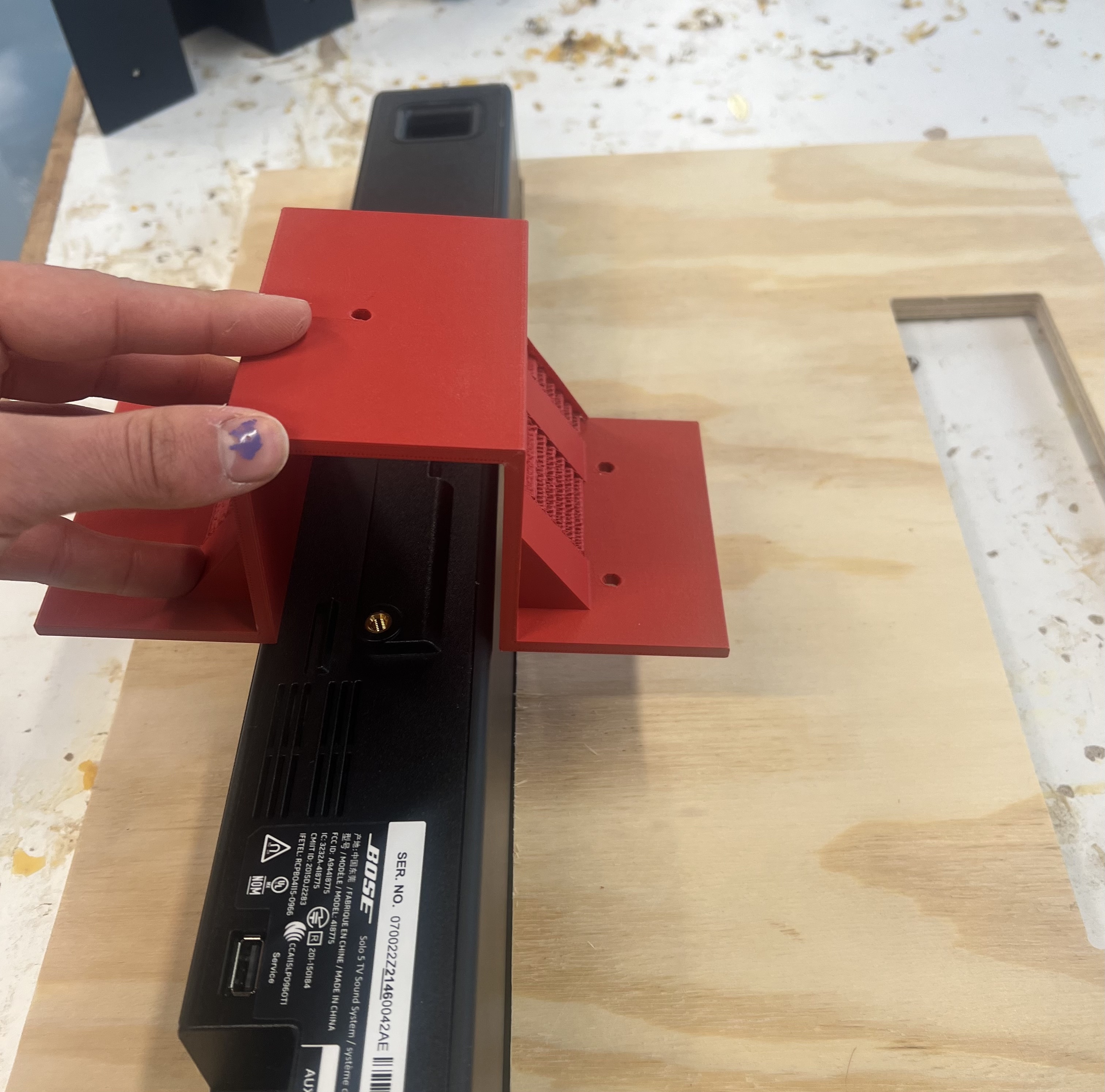}
  \caption{An early version of the 3D printed speaker brackets.}
  \label{fig:pcb}
\end{figure}

The purpose of using lights as opposed to some sort of LCD screen with symbols as in prior work was to enable the performer to communicate their performance visually as well as aurally, without requiring prior knowledge of tuning or microtonal music theory. Although it is underutilized in the current version of LIMITER, the individually addressable marquee lights can also change to show the color that is currently being drawn with. The sound synthesis engine uses the Teensy Audio Library\footnote{\url{https://www.pjrc.com/teensy/gui/}}, a patch-based audio system similar to other graphic programming languages such as MaxMSP or PureData.

\section{Results}

While the author had composed and performed with LIMITER extensively during its development, its first major public exhibition came during December of 2024, during a \anon{MIT Media Lab} community open house. During this event, the author had the opportunity to play for an audience of over one hundred people. Following the performance, a number of audience members lined up to play the interface, thus proving its desired dual function as an expressive musical instrument and a game for people to enjoy. 

Additionally, in February of 2025, a software version of LIMITER for use with a USB joystick was developed as part of a \anon{Media Lab} educational initiative and collaboration with the \anon{Dallas} Symphony Orchestra. This version will be deployed and used by local schools and children as a tool for learning about microtonality and just intonation and will provide further opportunities for refining the interface.

\section{Discussion}

The following section aims to discuss the effectiveness of LIMITER through various paradigms for evaluating new digital musical instruments. 

\subsection{Gamified Interfaces vs. Traditional DMIs}

The "player/controller/sound producing object" framework proposed first by Perry Cook \cite{cookremut} has long been used by the NIME community to consider new DMIs \cite{xiaokine}. As Cook argues, the natural flaw of these systems that involve constructing an artificial controller for some non-acoustically generated sound is three-fold: (1) the lack of haptic feedback from the controller/instrument to the player, (2) the presence of distortions/delays, and (3) the absence of any notion that the sound comes from the instrument itself \cite{cookremut}. These flaws largely explain the lack of success of many electronic instruments compared to traditional acoustic ones where the physical response of playing the instrument is intimately connected to the actual sonic output.

LIMITER, despite falling into this "player/controller/sound producing object" framework has a number of characteristics that enable it to escape the aforementioned flaws of such interfaces and achieve a musical experience that feels responsive, intuitive, and expressive for the performer and audience. LIMITER has very little latency, and the gestures and mappings are both predictable and learnable, while leaving room for creative compositional decisions. For (1), LIMITER leverages the control mechanism  of the arcade joystick and buttons, a relationship that has been shown to be satisfying precisely because of the inherently physical properties of the interface; while not literally "haptic" in the modern sense, an arcade joystick nonetheless exhibits the same properties of physical feedback that has been shown to increase a user's quality of experience in other modalities \cite{haptics}.

While LIMITER does primarily use digital sound generation, its status as both  an instrument and an arcade machine emphasizes the visceral sonic relationship afforded by its physical shape. Arcade machines have achieved such great popularity in part because of their uniquely multimodal and multisensory playing experience, causing the very notion of the "arcade aesthetic" to become entrenched in  popular culture as a whole \cite{culturecoin}. LIMITER, similarly, channels this by presenting the bright, colorful lights of the marquee and screen, the lightly resistant joystick and buttons, and of course the enormous spectacle of approaching a massive sound-producing body, thus emphasizing its physical form to the user. These factors help counter (3) that would otherwise make performance seem disembodied and therefore, unsatisfying.  

\subsection{Games as Musical Instruments}

The decision to construct LIMITER as an arcade machine as opposed to any number of other natural physical forms it could have taken arose as a result of the author's desire to take the unintuitive and esoteric system of just intonation and make it accessible by placing it in a familiar interface. As one of Perry Cook's Principles for Designing Computer Music Controllers states, "Everyday objects suggest amusing controllers" \cite{perryprinciples}. While an arcade machine is not an "an everyday object" in the conventional sense, LIMITER represents a leap forward in terms of transforming an object that has already a strong nostalgic presence in the minds of its players into a new conduit for digital sound production.

LIMITER's status as a game and a musical interface makes it significantly easier to engage with. Designer and Stanford University Professor Ge Wang humorously titles this the "tofu burger principle" \cite{geartful} - it is a gentle introduction to a, in this case, largely unfamiliar sound world. In short, LIMITER encourages engagement with what would otherwise feel like an overly academic, über-serious element of contemporary classical music.

\begin{figure}
  \centering
  \includegraphics[width=0.44\textwidth]{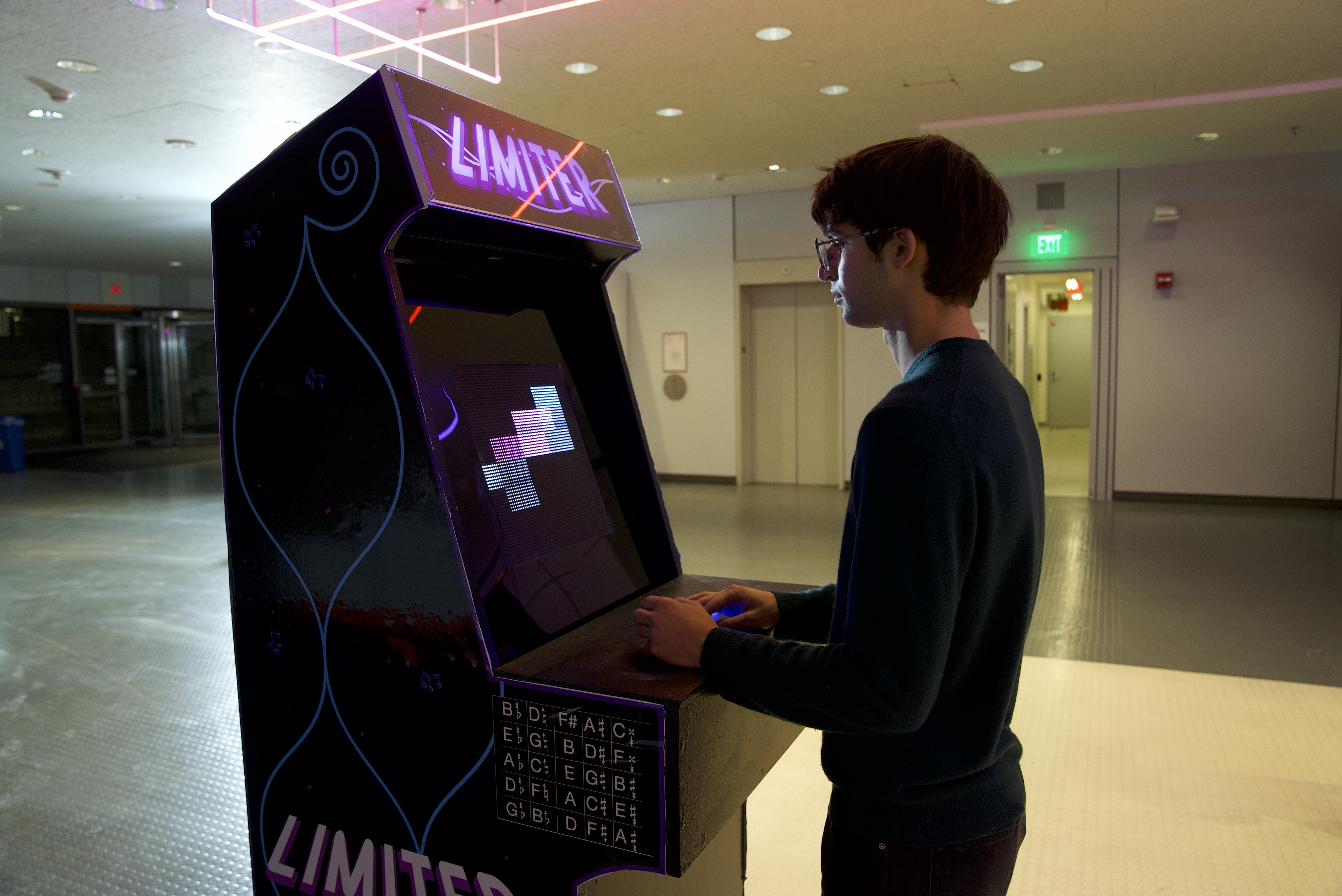}
  \label{fig:marquee}
\end{figure}

\section{Future Work}

There are a number of ways in which LIMITER can be improved for future installations and performances. Most importantly, the author aims to gather formal, large-scale user feedback from a range of musical backgrounds to examine the potential creativity-enhancing effects of the interface. Additionally, from a musical perspective, the author believes that there is further room to develop the sound system by adding additional oscillators,  audio effects, and other complexities to create even more varied types of expression. Finally, the instrument can be made more intuitive, which would enable its deployment in environments like actual arcades and other public places where passersby with no mediator or instruction set could stop and easily perform with it. 

\section{Conclusion}

In this paper we have introduced LIMITER, a gamified musical instrument for just intonation that enhances creative exploration through a playful and intuitive visual interface. We have also outlined the physical makeup and assembly of the instrument, as well as the various electronic systems that enable its function. LIMITER allows for the user to compose and perform music in a novel way, giving the ability for new, expressive musical experiences to both amateur and professional musicians alike.





\begin{acks}

  This project was made possible with the incredible support and mentorship of Professor Tod Machover and the Opera of the Future research group at the MIT Media Lab. The author would additionally like to thank Professor Paul Liang and the Multisensory Intelligence group for their generous funding of the project and Professor Neil Gershenfeld for his assistance with the physical fabrication. 

\end{acks}

\section{Ethical Standards}

This project follows the standards of the NIME ethical code of conduct. As there was no formal user study, no human participants were engaged or recruited. The author recognizes that certain fabrication processes used contribute in some way to environmental damage.


\bibliographystyle{ACM-Reference-Format}
\bibliography{sample-references}

\end{document}